\documentclass[sigconf]{acmart}
\AtBeginDocument{%
  }

\setcopyright{acmlicensed}
\copyrightyear{2025}
\acmYear{2025}
\acmDOI{XXXXXXX.XXXXXXX}

\acmConference[EASE 2025]{The 29th International Conference on Evaluation and Assessment in Software Engineering}{17–20 June, 2025}{Istanbul, Türkiye}

\acmISBN{978-1-4503-XXXX-X/2018/06}

\begin{document}

\title{Comparative Analysis of AI-Driven Security Approaches in DevSecOps: Challenges, Solutions, and Future Directions}

\author{Farid Binbeshr}
\email{farid.binbeshr@kfupm.edu.sa}
\orcid{1234-5678-9012}
\affiliation{%
  \institution{Interdisciplinary Research Center for Intelligent Secure Systems, King Fahd University of Petroleum and Minerals}
  \city{Dhahran}
  \country{Saudi Arabia}
}

\author{Muhammad Imam}
\email{mimam@kfupm.edu.sa}
\orcid{0000-0001-9131-6964}
\affiliation{%
  \institution{Computer Engineering Department and Interdisciplinary Research Center for Intelligent Secure Systems, King Fahd University of Petroleum and Minerals}
  \city{Dhahran}
  \country{Saudi Arabia}
}
% \affiliation{%
%   \institution{Interdisciplinary Research Center for Intelligent Secure Systems, King Fahd University of Petroleum and Minerals}
%   \city{Dhahran}
%   \country{Saudi Arabia}
% }

%%

\renewcommand{\shortauthors}{Faid Binbeshr and Muhammad Imam.}

\begin{abstract}
The integration of security within DevOps, known as DevSecOps, has gained traction in modern software development to address security vulnerabilities while maintaining agility. 
Artificial Intelligence (AI) and Machine Learning (ML) have been increasingly leveraged to enhance security automation, threat detection, and compliance enforcement. 
However, existing studies primarily focus on individual aspects of AI-driven security in DevSecOps, lacking a structured comparison of methodologies. 
This study conducts a systematic literature review (SLR) to analyze and compare AI-driven security solutions in DevSecOps, evaluating their technical capabilities, implementation challenges, and operational impacts. The findings reveal gaps in empirical validation, scalability, and integration of AI in security automation. The study highlights best practices, identifies research gaps, and proposes future directions for optimizing AI-based security frameworks in DevSecOps.
\end{abstract}





\maketitle

%%%%%%%%%%%%%%%%%%%%%%%%%%%%%%%%%%%%%%%%%%%%%%%%%%%%%%%%%%%%%%%%%%%%%%
\section{Introduction}
%%%%%%%%%%%%%%%%%%%%%%%%%%%%%%%%%%%%%%%%%%%%%%%%%%%%%%%%%%%%%%%%%%%%%%
% \cite{S1, S2, S3, S4, S5, S6, S7, S8, S9, S10, S11, S12, S13, S14, S15, S16, S17, S18}
The integration of security into the DevOps lifecycle, commonly known as DevSecOps, is essential for addressing security risks in modern software development. 
DevOps has revolutionized software engineering by emphasizing automation, continuous integration (CI), and continuous delivery (CD). 
However, the rapid pace of deployment often leads to security vulnerabilities due to inadequate security controls. 
DevSecOps seeks to embed security practices into every stage of the software development lifecycle (SDLC), ensuring proactive threat mitigation. 
Despite its growing adoption, the effectiveness of various DevSecOps frameworks remains an area of active research \cite{pakalapati2023unlocking}. 
Existing studies highlight the potential of Artificial Intelligence (AI) and Machine Learning (ML) in security automation but lack a structured comparison of different frameworks, leaving a gap in understanding the best practices for implementation \cite{schieseck2024formal, fu2024ai}.

Several studies have explored AI-driven security in DevSecOps, yet a structured comparison of these methodologies remains limited. 
\citet{rajapakse2022challenges} identified challenges in security automation and tool integration but did not evaluate proposed solutions. 
\citet{prates2019devsecops} introduced DevSecOps metrics but lacked a standardized framework for assessing security effectiveness. 
AI-driven threat detection has been studied in IoT security by \citet{bahaa2021monitoring}, though its broader applicability to DevSecOps remains unclear. 
Similarly, \citet{mboweni2022systematic} highlighted gaps in ML integration within DevOps, particularly in security automation. 
Compliance remains a challenge, with \citet{rajapakse2022challenges} also noting the lack of consensus on shift-left security practices. 
Other studies \cite{fu2024ai, camacho2024unlocking} on AI-driven security in DevSecOps reveal significant gaps, such as the limited exploration of specific security tasks and the absence of comprehensive evaluation methods. 
While these studies provide valuable insights into individual aspects of DevSecOps, a comparative analysis of existing frameworks remains missing. 
This study aims to fill this gap by systematically reviewing and evaluating DevSecOps approaches, focusing on their effectiveness in modern software development.

The primary aim of this study is to evaluate the effectiveness of AI-driven techniques in DevSecOps.
To achieve this aim, a systematic literature review (SLR) methodology was employed to identify current AI and ML solutions designed for DevSecOps.
Then, a comparative analysis of these solutions was conducted with respect to their technical capability, implementation requirement, and operational impact.
This study contributes to the field by:
\begin{itemize}
    \item Identifying existing AI/ML solutions desgined for DevSecOps.
    \item Providing a structured comparison of AI-driven security solutions within DevSecOps.
    \item Identifying key implementation challenges and best practices for security automation.
    \item Highlighting gaps in existing research and proposing future research directions.
\end{itemize}

The remainder of this paper is structured as follows: Section 2 details the methodology, including data sources and selection criteria. 
Section 3 presents the results of the SLR. 
Section 4 provides the comparative analysis, emphasizing key security frameworks and AI-driven approaches. 
Section 5 discusses implementation challenges, operational impacts, key findings, and research implications. 
Additionally, it outlines future research directions. 
Finally, Section 6 concludes the study, summarizing its contributions and providing recommendations.

%%%%%%%%%%%%%%%%%%%%%%%%%%%%%%%%%%%%%%%%%%%%%%%%%%%%%%%%%%%%%%%%%%%%%%
\section{Methodology}
%%%%%%%%%%%%%%%%%%%%%%%%%%%%%%%%%%%%%%%%%%%%%%%%%%%%%%%%%%%%%%%%%%%%%%
\subsection{Database Selection and Search Strategy}

To conduct our SLR on the intersection of \textit{SecDevOps, DevOpsSec, and SecOps} with \textit{Machine Learning, Deep Learning, and Artificial Intelligence}, we carefully selected reputable academic databases. 
The selection criteria were based on their relevance to cybersecurity, software engineering, and artificial intelligence research, as well as their inclusion of peer-reviewed journal articles and conference proceedings. 
The objective was to ensure comprehensive coverage of studies that address the research aim posed in this study.
The following databases were chosen for the search: \textit{ACM Digital Library, IEEE Xplore, Scopus, Springer Link, Web of Science, and ScienceDirect}. 
These databases provide extensive coverage of computing, security, and artificial intelligence research, making them well-suited for this study.

While both SLRs and Mapping Literature Reviews (MLRs) are recognized secondary research methodologies, an SLR was preferred for this study due to its emphasis on critical appraisal, synthesis, and reproducibility. 
Unlike MLRs, which focus on broad categorization and coverage, SLRs are better suited to evaluating specific research questions, identifying evidence-based practices, and deriving analytical insights.
Given the rapidly evolving and technically nuanced nature of AI applications in DevSecOps, an SLR provides a structured and rigorous mechanism to assess quality, extract comparative features, and produce actionable insights for researchers and practitioners alike.

\subsection{Search String and Refinement}

To maintain consistency in literature retrieval, a standardized search string was employed across all databases. 
The primary search string used in the retrieval process was formulated as follows:

\begin{quote}
(SecDevOps OR DevOpsSec OR SecOps OR (secur* AND DevOps)) AND ("machine learning" OR "deep learning" OR "artificial intelligence" OR Intelligent)
\end{quote}

Given the variations in search functionalities across different databases, refinement strategies were applied to enhance the relevance of retrieved studies and exclude unrelated research. The refinements were implemented as follows:

\begin{itemize}
    \item \textbf{ACM Digital Library} – The search was restricted to conference proceedings, yielding two relevant results.
    \item \textbf{IEEE Xplore} – The search included both conference papers and journal articles, refining the results from an initial count of 57 to 52.
    \item \textbf{Scopus} – Review papers were excluded to retain only primary research articles, reducing the number of selected studies from 82 to 56.
    \item \textbf{Springer Link} – The original search retrieved several irrelevant studies. To improve precision, the query was modified to emphasize logical keyword connections, reducing the results from 10 to 2.
    \item \textbf{Web of Science} – Minimal refinement was necessary, with 21 out of 22 retrieved papers being retained.
    \item \textbf{ScienceDirect} – The search was refined to title, abstract, and author-specified keywords, yielding 11 relevant articles.
\end{itemize}

In addition to the aforementioned databases, other sources such as \textbf{Wiley} and \textbf{Google Scholar} were initially considered but later excluded due to the irrelevance of retrieved articles or insufficient results. 
Other databases, including \textbf{Microsoft Academic, CiteSeerX, IOPscience, and Taylor \& Francis}, were evaluated but not included in the final selection due to a lack of relevant publications.

\subsection{Inclusion Criteria}

To ensure that the selected studies are relevant and contribute effectively to this SLR, predefined inclusion criteria were applied. 
The inclusion criteria are as follows:

\begin{itemize}
    \item \textbf{IC1: Articles that employ AI in DevSecOps} – The study must explore the use of AI in DevSecOps to ensure its relevance to the research aim.
    \item \textbf{IC2: Primary research articles} – Only primary research studies were considered, as SLRs focus on analyzing original research contributions.
    \item \textbf{IC3: Articles published in English} – Studies must be in English, as translating papers from other languages may introduce errors or misinterpretations.
    \item \textbf{IC4: Articles with full access} – Only studies with complete access were included to allow thorough review and analysis.
\end{itemize}

% By applying these criteria, we ensured that the selected papers contribute meaningful insights to the study while maintaining clarity and accessibility.

\subsection{Study Selection}
The study selection followed a structured process in multiple stages. 
First, the search string was applied to the selected databases, and the retrieved results were imported into EndNote for management. 
Duplicate records were removed, and the remaining studies were screened based on their titles and abstracts. 
Full-text articles were then reviewed to ensure they met the inclusion criteria, and relevant data items were extracted. 
The selection process was conducted by two authors to ensure accuracy and consistency.

\subsection{Quality Assessment}
\label{sec: quality assessment}
A quality assessment checklist was developed to evaluate the relevance and reliability of the selected studies. 
The checklist consisted of nine criteria designed to assess the contribution of each study to this review. 
These criteria were based on established guidelines, including the CASP Qualitative Checklist~\cite{casp2015casp} and Keele et al.’s~\cite{keele2007guidelines} systematic review guidelines for software engineering. 
Each study was assigned a quality score per criterion, with 1 for "fully met," 0.5 for "partially met," and 0 for "not met." 
The total score ranged from 0 to 9, where higher scores indicated higher quality.

\subsection{Data Extraction Strategy}

To systematically collect relevant information from the selected studies, we used an Excel spreadsheet to record key data items essential for answering the research questions and assessing study quality. 
The extracted data included:

\begin{itemize}
    \item \textbf{Bibliographic Information:} Study ID, title, year, author(s), author type, publication type, and venue name.
    \item \textbf{Research Focus:} Identified problems, proposed solutions, and study context or application.
    \item \textbf{AI Approach:} Techniques, models, or algorithms used, along with dataset details.
    \item \textbf{Methodology:} Data collection methods, data analysis techniques, evaluation metrics, comparison methods, and performance results.
    \item \textbf{Findings and Challenges:} Key findings, identified challenges, opportunities, and study limitations.
\end{itemize}

%%%%%%%%%%%%%%%%%%%%%%%%%%%%%%%%%%%%%%%%%%%%%%%%%%%%%%%%%%%%%%%%%%%%%%
\section{Results}
%%%%%%%%%%%%%%%%%%%%%%%%%%%%%%%%%%%%%%%%%%%%%%%%%%%%%%%%%%%%%%%%%%%%%%
\subsection{Study Selection Results}
\begin{figure}[ht!]
    \centering
    \includegraphics[width=0.475\textwidth]{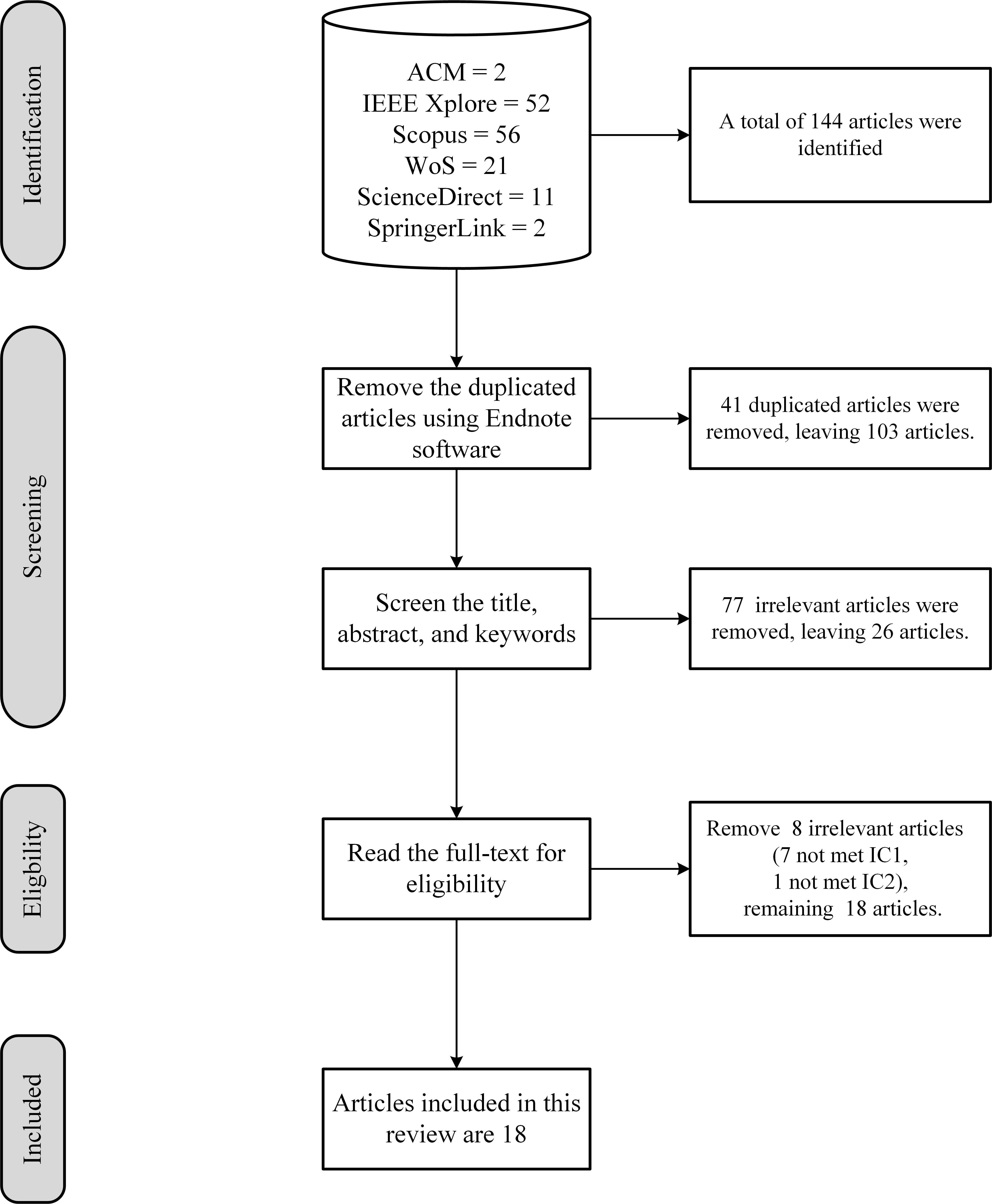}
    \caption{Flowchart showing the systematic selection process for the SLR. Initially, 144 articles were identified from six databases. After removing duplicates and irrelevant studies, 18 articles were included in the final review.}
    \label{fig:study-flow}
\end{figure}

The study selection process followed a systematic approach, as illustrated in Figure~\ref{fig:study-flow}. 
Initially, 144 articles were identified from six databases: ACM (2), IEEE Xplore (52), Scopus (56), Web of Science (21), ScienceDirect (11), and SpringerLink (2). 
After removing 41 duplicate articles using Endnote software, 103 unique articles remained. 
A screening of titles, abstracts, and keywords resulted in the exclusion of 77 irrelevant articles, leaving 26 for further evaluation. 
Full-text analysis was conducted to assess eligibility, leading to the removal of eight additional articles—seven for not meeting Inclusion Criterion 1 (IC1) and one for not meeting Inclusion Criterion 2 (IC2). Ultimately, 18 articles were included in the final review.

\subsection{Study Characteristics}
Table~\ref{tab:publications} presents a structured overview of selected publications on DevSecOps and security practices, categorizing them by study ID, author, publication year, type, and venue. 
The dataset spans from 2015 to 2024, reflecting the growing academic interest in DevSecOps methodologies and security frameworks. 
The majority of studies are recent, particularly from 2022 onwards, indicating an increasing focus on integrating security into software development pipelines. 
Journals and conferences contribute equally to the dissemination of DevSecOps research, each representing 50\% of the selected publications. 
Notable journal venues include \textit{Computers \& Security, The Journal of Systems and Software, and IEEE Transactions on Parallel and Distributed Systems}, while key conferences include \textit{ACM/SPEC International Conference on Performance Engineering, IEEE International Conference on Software Analysis, Evolution and Reengineering (SANER), and CLOSER}.

\begin{small}
    \begin{table*}[h!t]
       \caption{Overview of the 18 selected publications, categorized by ID, author, year, type, and venue. The dataset spans from 2015 to 2024, with most studies published from 2022 onwards, reflecting increasing academic interest in DevSecOps methodologies. Notable publication venues include IEEE Transactions, ACM conferences, and journals such as Computers \& Security.}
        \label{tab:publications}
        \centering
        \small
        \begin{tabular}{ccccp{8cm}}
        \toprule
        ID & Author (Year) & Reference & Type & Venue \\
        \midrule
        S1 & \citeauthor{S1} (\citeyear{S1}) & \cite{S1} & Journal & International Journal of Intelligent Systems and Applications in Engineering (IJISAE) \\
        S2 & \citeauthor{S2} (\citeyear{S2}) & \cite{S2} & Journal Article & Heliyon \\
        S3 & \citeauthor{S3} (\citeyear{S3}) & \cite{S3} & Journal Article & IEEE Communications Magazine \\
        S4 & \citeauthor{S4} (\citeyear{S4}) & \cite{S4} & Journal Article & Computers \& Security \\
        S5 & \citeauthor{S5} (\citeyear{S5}) & \cite{S5} & Journal & Software Quality Journal \\
        S6 & \citeauthor{S6} (\citeyear{S6}) & \cite{S6} & Conference Paper & IEEE International Conference on Software Analysis, Evolution and Reengineering (SANER) \\
        S7 & \citeauthor{S7} (\citeyear{S7}) & \cite{S7} & Conference & ACM/SPEC International Conference on Performance Engineering (ICPE '23 Companion) \\
        S8 & \citeauthor{S8} (\citeyear{S8}) & \cite{S8} & Journal Article & Journal of Experimental \& Theoretical Artificial Intelligence \\
        S9 & \citeauthor{S9} (\citeyear{S9}) & \cite{S9} & Conference & 30th Telecommunications Forum TELFOR 2022 \\
        S10 & \citeauthor{S10} (\citeyear{S10}) & \cite{S10} & Conference Paper & 26th International Conference on Knowledge-Based and Intelligent Information \& Engineering Systems \\
        S11 & \citeauthor{S11} (\citeyear{S11}) & \cite{S11} & Journal Article & Journal of Cloud Computing \\
        S12 & \citeauthor{S12} (\citeyear{S12}) & \cite{S12} & Conference Paper & 2nd International Mobile, Intelligent, and Ubiquitous Computing Conference (MIUCC) \\
        S13 & \citeauthor{S13} (\citeyear{S13}) & \cite{S13} & Conference Paper & Proceedings of the International Conference on Intelligent Computing and Control Systems (ICICCS 2020) \\
        S14 & \citeauthor{S14} (\citeyear{S14}) & \cite{S14} & Conference & IEEE ESSCA 2020 - Workshop on Emerging Security Solutions for Critical Applications \\
        S15 & \citeauthor{S15} (\citeyear{S15}) & \cite{S15} & Journal & The Journal of Systems and Software \\
        S16 & \citeauthor{S16} (\citeyear{S16}) & \cite{S16} & Conference Paper & DAIS 2020, LNCS 12135, Springer Nature Switzerland AG 2020 \\
        S17 & \citeauthor{S17} (\citeyear{S17}) & \cite{S17} & Journal Article & IEEE Transactions on Parallel and Distributed Systems \\
        S18 & \citeauthor{S18} (\citeyear{S18}) & \cite{S18} & Conference & 5th International Conference on Cloud Computing and Services Science (CLOSER-2015) \\
        \bottomrule
        \end{tabular}
        \vspace{6pt}  % Add 6 points of vertical space here
    \end{table*}
\end{small}

Table~\ref{tab:ai_clusters} presents a structured categorization of AI models and techniques applied in studies. 
The table clusters AI techniques into five main groups based on their application domain and purpose. 
The DevOps \& Security Automation cluster includes AI-driven DevOps practices, AIOps, Infrastructure-as-Code, and automated security enforcement tools, with studies such as S1, S5, S11, and S16, focusing on securing DevOps pipelines and infrastructure. 
The Machine Learning \& Anomaly Detection cluster covers deep learning-based security approaches, including Multi-layer Perceptron (MLP), Random Forest (RF), and LSTM networks for detecting anomalies in system logs and security operations, as seen in S2, S8, S17, and others. 
The Security Testing \& Vulnerability Assessment group encompasses methodologies for security testing and vulnerability detection, such as model-based security testing and static code analysis, applied in S4, S7, S12, and S14. 
The Threat Detection \& Risk Assessment cluster integrates AI-driven threat intelligence, risk authentication models, and attack-defense strategies to enhance security, with relevant studies including S3, S9, S10, and S15. 
Lastly, the Cloud \& Multi-Cloud Security category includes AI techniques designed for securing cloud-based applications and multi-cloud environments, as demonstrated in S6, S13, and S18. 
This structured classification provides a clear overview of how AI is leveraged in different security domains.

\begin{small}
    \begin{table*}[h!t]
       \caption{Categorization of AI models and techniques into five clusters—DevOps \& Security Automation, Machine Learning \& Anomaly Detection, Security Testing \& Vulnerability Assessment, Threat Detection \& Risk Assessment, and Cloud \& Multi-Cloud Security—to provide a structured overview of how AI is leveraged across different security domains in DevSecOps.}
        \label{tab:ai_clusters}
        \centering
        \small
        \begin{tabular}{p{3.5cm}p{4.5cm}p{5cm}p{3cm}}
        \toprule
        Cluster & Techniques/Models & Purpose & Studies \\
        \midrule
        DevOps \& Security Automation & AI-driven DevOps, AIOps, IaC, Continuous Security Monitoring, Terraform, Ansible & Automating security enforcement and compliance in DevOps & S1, S5, S11, S16 \\
        Machine Learning \& Anomaly Detection & MLP, RF, LSTM, DCNN, DBN with FAE-GWO, Unsupervised and Supervised Learning & Identifying security anomalies, intrusion detection, and threat prediction & S2, S8, S17 \\
        Security Testing \& Vulnerability Assessment & Model-based security testing, Static Code Analysis, DSCA, DroidAutoML, Feature Extraction & Automating security testing, vulnerability detection, and code analysis & S4, S7, S12, S14 \\
        Threat Detection \& Risk Assessment & Risk Authentication Models, Attack-Defense Trees, AI-based Threat Intelligence & Identifying and mitigating security risks dynamically & S3, S9, S10, S15 \\
        Cloud \& Multi-Cloud Security & µDetector for Kubernetes, MUSA framework for security-by-design in multi-cloud & Securing cloud-based applications and multi-cloud environments & S6, S13, S18 \\
        \bottomrule
        \end{tabular}
        \vspace{6pt}  % Add 6 points of vertical space here
    \end{table*}
\end{small}

\subsection{Quality Assessment Results}
\begin{figure}[h]
    \centering
    \includegraphics[width=0.5\textwidth]{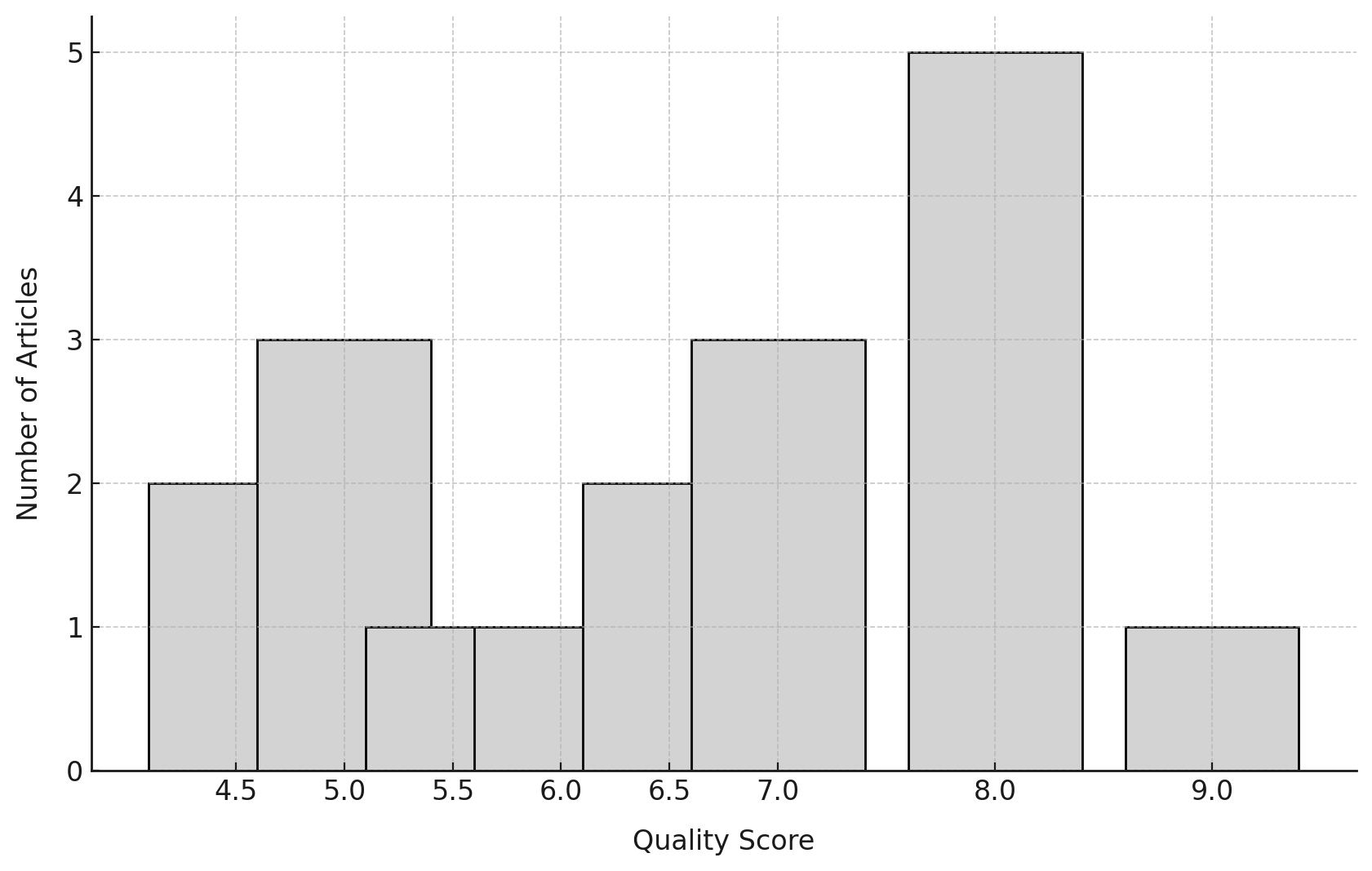}
    \caption{Bar chart displaying the quality scores of the 18 selected studies, assessed using a predefined checklist with scores ranging from 0 to 9. Most studies scored between 4.5 and 9, reflecting high methodological quality. However, some studies lacked detailed explanations of limitations, highlighting areas for improvement in reporting transparency.}
    \label{fig:qa_distribution}
\end{figure}

The quality assessment of the selected studies was conducted based on nine predefined criteria, evaluating different aspects such as study design, data collection, analysis, and conclusions. 
Each criterion was scored as "fully met" (1 point), "partially met" (0.5 points), or "not met" (0 points), as described in Section~\ref{sec: quality assessment}.
The total quality score for each study was computed by summing these scores.
Figure~\ref{fig:qa_distribution} presents the distribution of quality scores across the studies. 
The majority of studies scored between 4.5 and 9, indicating a generally high methodological quality among the included works. 
A significant proportion of studies attained scores of 7.0 or above, demonstrating well-documented research methods and findings.

The quality assessment revealed some variations in reporting across studies. 
While most studies provided comprehensive details on their research aim, methodology, and data analysis, some lacked thorough explanations of study settings and limitations. 
These inconsistencies highlight areas where improvements in methodological transparency and reporting could be made. 
Overall, the assessment suggests that the included studies provide a strong foundation for the study, with most meeting the key quality criteria to a satisfactory degree.

Having identified the key studies and their characteristics, the next section provides a detailed comparative analysis of the AI-driven security approaches, focusing on their technical capabilities, implementation challenges, and operational impacts.

%%%%%%%%%%%%%%%%%%%%%%%%%%%%%%%%%%%%%%%%%%%%%%%%%%%%%%%%%

\section{Comparative Analysis}
\label{Sec:ComparativeAnalysis}
The widespread adoption of DevOps methodologies has significantly transformed software development by prioritizing automation, continuous integration, and team collaboration. 
As noted in the introduction, integrating security into DevOps is challenging because organizations must balance agility with stringent security practices \cite{S1, S2}. 

This study synthesizes findings from 18 selected references to evaluate security frameworks, automation tools, implementation challenges, and operational impacts in DevSecOps environments.
The analysis is structured across three key dimensions:

\begin{enumerate}
    \item \textbf{Technical Capability Analysis} – Evaluates security frameworks, automation tools, and privacy-enhancing techniques.
    \item \textbf{Implementation Requirements Analysis} – Examines computational overhead, tool compatibility, and scalability.
    \item \textbf{Operational Impact Assessment} – Assesses the effects on DevOps agility, developer productivity, and risk management.
\end{enumerate}

\subsection{Technical Capability Analysis}
\label{Sec:TechnicalCapabilityAnalysis}

Integrating security frameworks into DevOps effectively mitigates risks in CI/CD pipelines. Studies S3, S4, and S8 underline the necessity of "shift-left" security, embedding security early in software development to address vulnerabilities proactively.

Threat modeling techniques, such as those in S4 and S10, connect vulnerability identification directly with testing strategies. The Attack-Defense Trees (ADTs) method (S10) systematically identifies vulnerabilities and verifies protective measures. Automated detection approaches, including SAST and DAST (S7), facilitate continuous vulnerability scanning. For instance, µDetector (S6) enhances Kubernetes security through anomaly detection in system calls.

Privacy frameworks, notably DevPrivOps (S3), integrate automated privacy risk assessments to strengthen privacy in DevSecOps. NLP techniques using LSTM networks (S11, S14) improve anomaly detection in logs and tracing data, enhancing security monitoring.

AI and ML automation significantly advance DevSecOps security. Studies S1 and S2 highlight automated remediation of vulnerabilities through AI-driven detection. Deep learning models (DBN, DCNN) effectively detect IoT threats in real-time (S8, S13). IaC tools like Terraform and Ansible (S9, S12) enforce automated security policies but face multi-cloud limitations due to inconsistent configurations (S9).

Compliance and monitoring automation, covered in S1 and S12, streamline regulatory compliance but face scalability challenges for ML-driven security monitoring (S6, S13). CI/CD pipeline security benefits from early security integration, although tools like CyberDevOps (S5) encounter high false-positive rates. Compatibility among security tools and inconsistencies in IaC checks (S7, S9) remain significant challenges.

\begin{figure*}[htbp]
\centering
\includegraphics[width=0.5\textwidth]{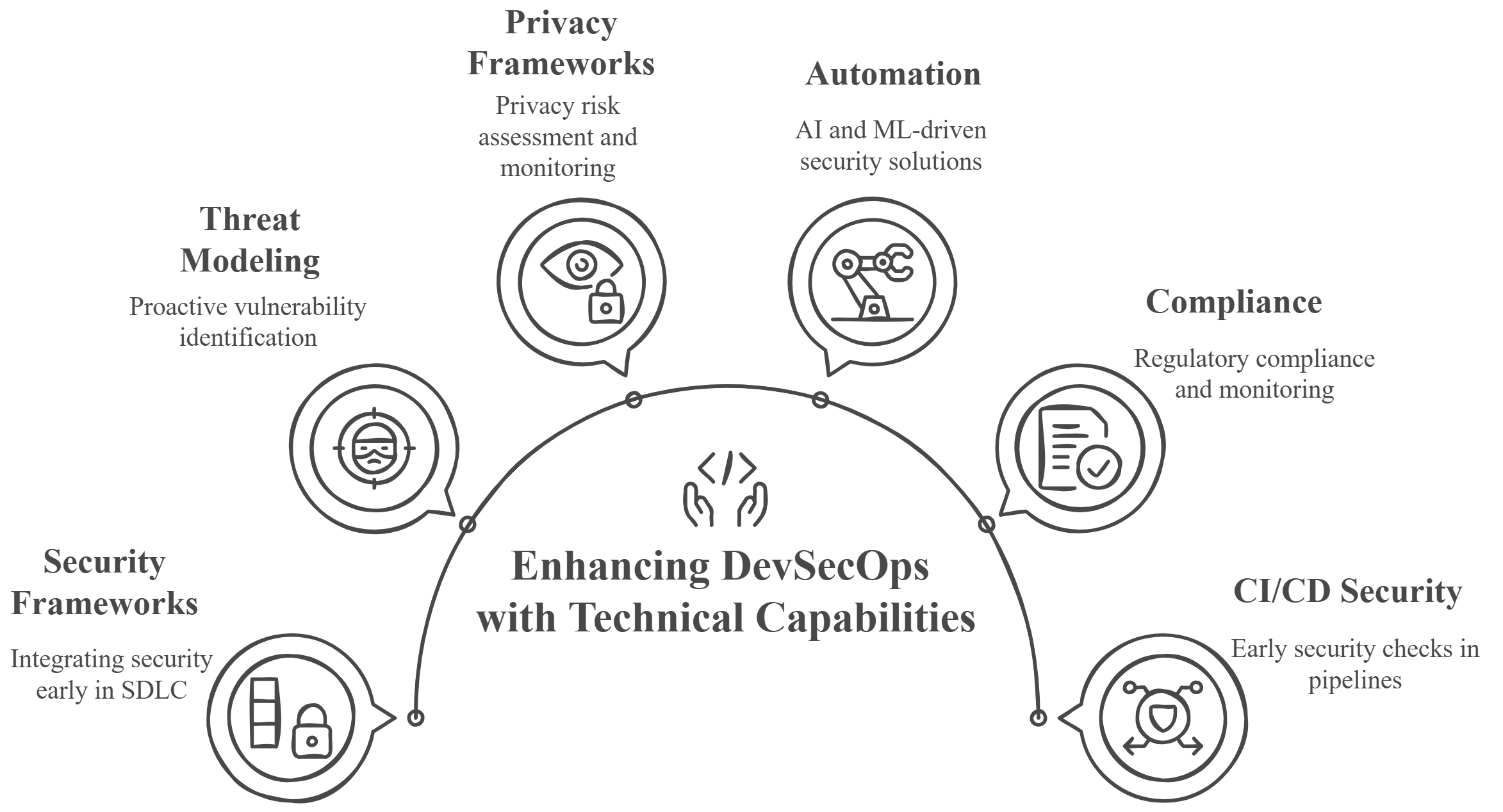}
\caption{Technical capabilities enhancing DevSecOps, covering key areas including security frameworks, threat modeling, privacy frameworks, automation, compliance, and CI/CD security.}
\label{fig:TechnicalCapabilitiesDevSecOps}
\end{figure*}

Overall, addressing these technical integration challenges through targeted research can significantly enhance DevSecOps security capabilities. 
Figure~\ref{fig:TechnicalCapabilitiesDevSecOps} visually summarizes these technical capabilities and their interactions within DevSecOps.

\subsection{Implementation Requirements Analysis}

Integrating ML-based security into DevOps introduces several implementation challenges. Computational overhead significantly limits real-time application and scalability. 
Deep Learning models (S8, S11, S13) require considerable resources, complicating their use for real-time threat detection in IoT and microservices. 
Additionally, privacy frameworks, such as those in S3, demand extensive computational capabilities, restricting large-scale deployments.

Tool compatibility issues also pose significant challenges.
For example, the Python-based IaC Analyzer (S9) lacks support for multiple Infrastructure-as-Code (IaC) formats, limiting interoperability. 
Similarly, dependency management problems complicate integrating frameworks like LOMOS (S7) into existing DevOps workflows.

Scalability and cloud-native security further complicate implementation. 
The µDetector framework (S6) can process significant volumes of system calls but faces latency at peak times. 
NLP-based anomaly detection (S11) struggles with interpretability at scale. 
Cloud-native security tools like AWS automation (S12) lack multi-cloud compatibility, while frameworks such as MUSA (S18) have limited empirical validation, raising concerns about real-world applicability.

\begin{figure}[htbp]
    \centering
    \includegraphics[width=0.475\textwidth]{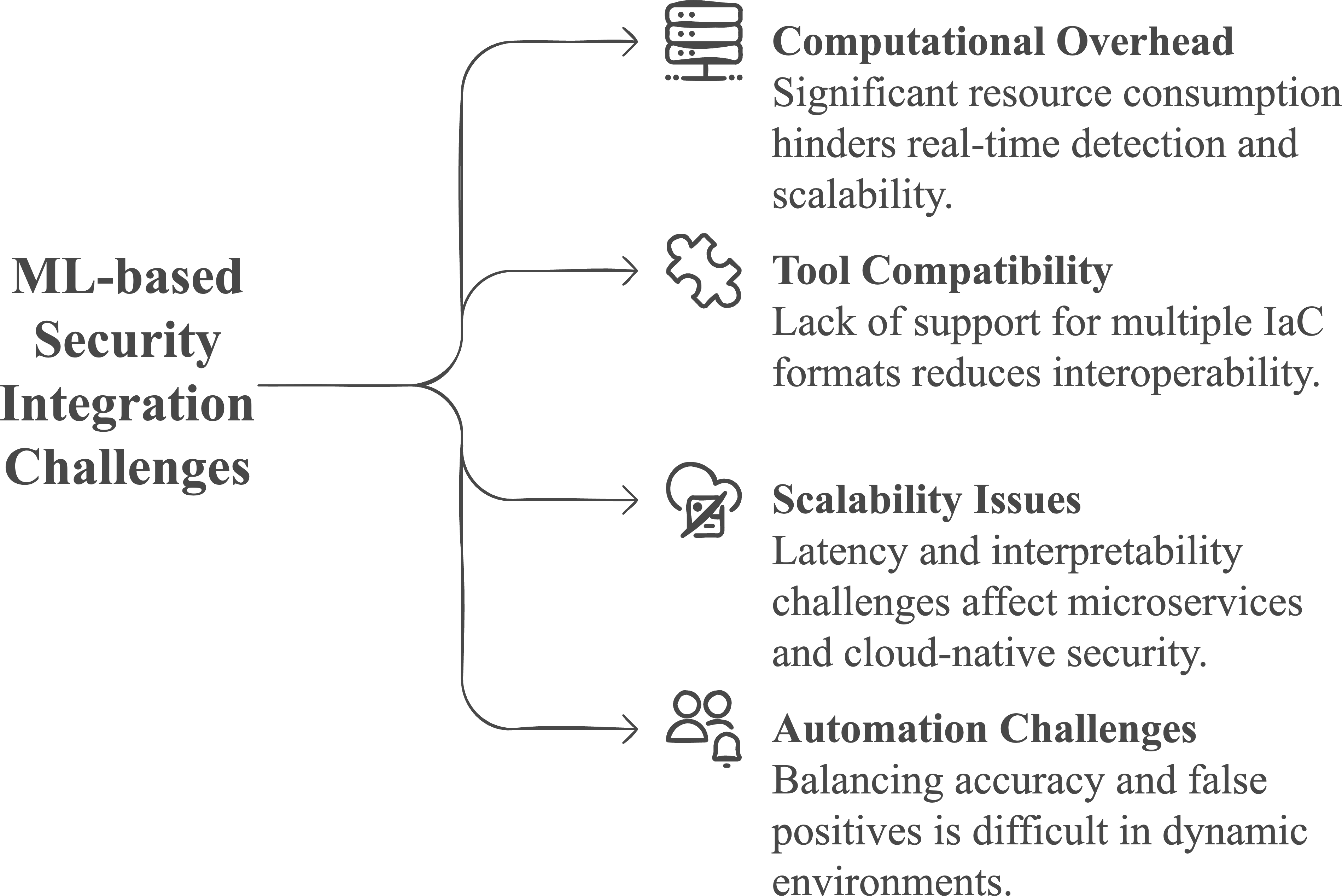}
    \caption{Overview of key challenges in integrating ML-based security into DevOps, including computational overhead, tool compatibility, scalability issues, and automation challenges.}
    \label{fig:MLsecurityIntegrationChallenges}
\end{figure}

Automation presents additional challenges, notably managing false positives and adapting to dynamic environments. 
CyberDevOps (S5) and LOMOS (S7) face difficulties balancing accuracy with the rate of false positives, often requiring manual intervention. 
Moreover, adaptive frameworks like the autotuning solution (S17) incur high computational costs, reducing efficiency in large deployments.
Figure~\ref{fig:MLsecurityIntegrationChallenges} summarizes these key implementation challenges.

\subsection{Operational Impact Assessment}
While automated security solutions improve resilience, they can also impact DevOps agility.
The security automation module in S12 decreases setup time by 83\%, leading to faster deployment.
Furthermore, S4 and S15 report that manual security assessments introduce delays in CI/CD pipelines, creating operational bottlenecks.

On the other hand, case studies from S1 and S12 indicate that AI-driven security enhances compliance in healthcare but lacks full visibility into cloud security risks.
AWS security automation helps mitigate SSH key exposure risks but does not provide integrated secret management capabilities.

When it comes to humans, the studies S5 and S15 highlight that developers often lack sufficient security expertise, emphasizing the need for security training programs.
As illustrated in S9, automated security tools minimize human errors but require real-time feedback mechanisms to support developers in securing code effectively.

\begin{figure*}[!th]
    \centering
    \includegraphics[width=0.5\textwidth]{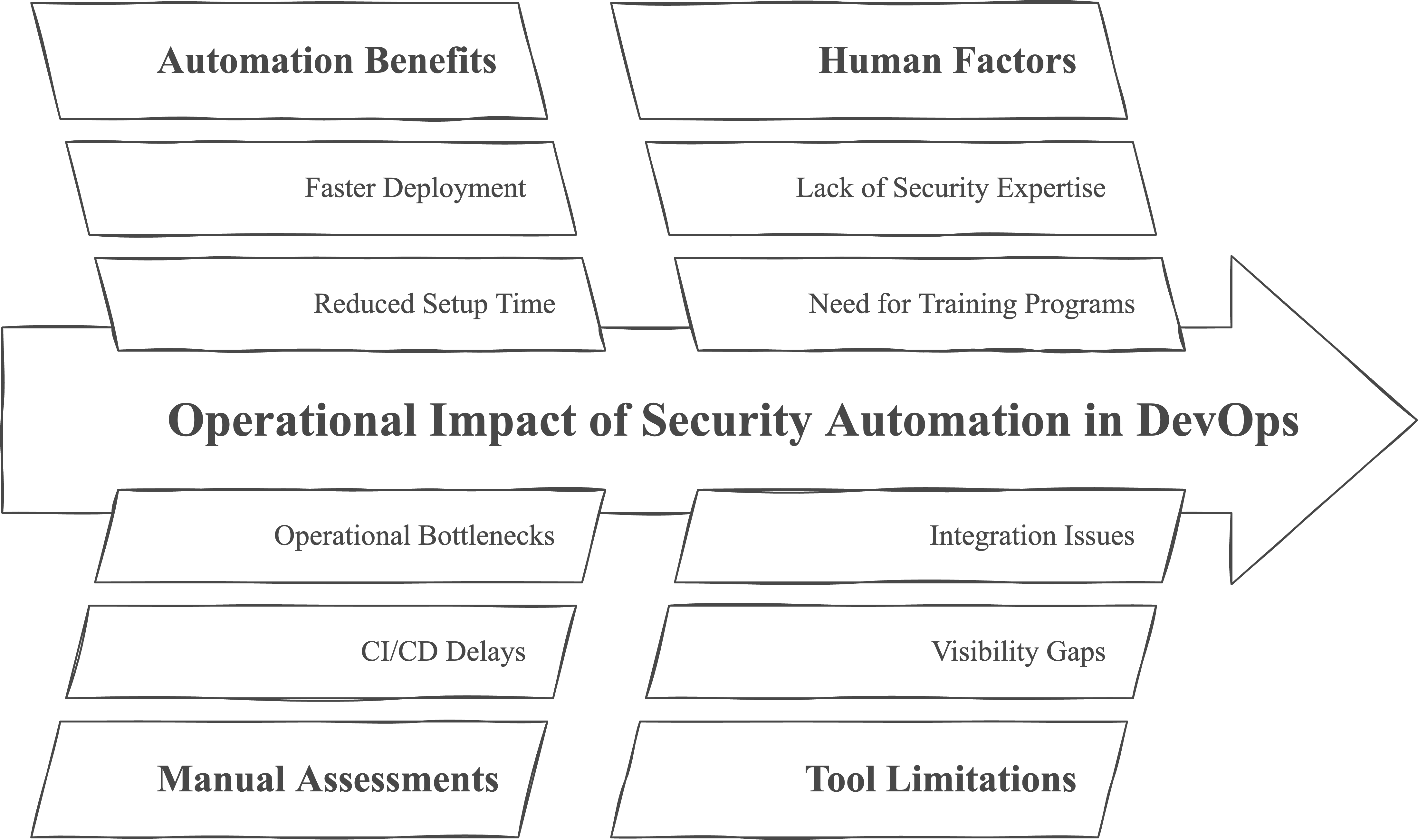}
    \caption{Diagram illustrating the benefits (e.g., faster deployment) and challenges (e.g., CI/CD delays) of security automation in DevOps, along with human factors like developer training needs.}
    \label{fig:OperationalImpact}
\end{figure*}

Figure~\ref{fig:OperationalImpact} summarizes key implementation challenges and their impact on security automation.

\subsection{Identified Gaps and Future Research Directions}
Despite increasing adoption of DevSecOps, several research gaps remain:

\begin{itemize}
    \item \textbf{Lack of Real-World Validation} – Many proposed solutions (S5, S18) lack empirical validation, making their practical effectiveness uncertain.
    \item \textbf{Privacy Beyond Compliance} – Current research focuses on regulatory compliance but overlooks aspects such as privacy quantification and user consent management (S3, S14).
    \item \textbf{Scalable Security Automation} – Existing ML-based security models require high computational resources, highlighting the need for lightweight and scalable alternatives (S6, S13).
    \item \textbf{Human-Centric Security} – Usability of security tools (S15) and developer-focused security training (S5) remain underexplored.
\end{itemize}

The comparative analysis of S1–S18 highlights the evolution of DevSecOps, emphasizing security automation, privacy concerns, and operational efficiency. While ML-based security solutions improve threat detection, high computational costs, tool fragmentation, and developer skill gaps remain challenges.

% %%%%%%%%%%%%%%%%%%%%%%%%%%%%%%%%%%%%%%%%%%%%%%%%%%%%%%%%

\subsection{Comparative Evaluation of AI-Driven DevSecOps Solutions}

Based on the analysis of the selected studies (S1--S18), as shown in Table~\ref{tab:comparative_table}, the AI-driven security approaches were grouped into five major clusters, and their performance was compared across key dimensions: detection rate, false positive rate, latency, and scalability.

\begin{table*}[h!t]
\centering
\caption{Clustered comparative evaluation of AI-driven DevSecOps solutions across key performance metrics, including detection rate, false positive rate, latency, and scalability. Solutions are grouped into five thematic categories based on technical focus and application domain, illustrating the trade-offs between detection performance, operational latency, computational overhead, and scalability.}
\label{tab:comparative_table}
\renewcommand{\arraystretch}{1.2}
\begin{tabular}{p{3.8cm}p{3.5cm}p{1.8cm}p{1.8cm}p{1.8cm}p{2.5cm}}
\toprule
\textbf{Category} & \textbf{Representative Studies} & \textbf{Detection Rate (\%)} & \textbf{False Positive Rate (\%)} & \textbf{Latency} & \textbf{Scalability} \\
\midrule
Microservice-focused \newline Threat Detection & \cite{S6,S7,S11} & 96.5--97.8\% & 1.8--3.2\% & Low & High \\
\midrule
Infrastructure-as-Code \newline and Policy-as-Code \newline Security & \cite{S1,S5,S9} & N/A \newline (configuration validation) & Moderate & Moderate & Medium \\
\midrule
Privacy- and Compliance- \newline Oriented Frameworks & \cite{S3,S4,S15} & N/A & N/A & High & Low--Medium \\
\midrule
IoT-Specific \newline Attack Detection & \cite{S8,S13,S14} & 95--98\% & 2--3\% & High & Low--Medium \\
\midrule
Multi-cloud and Cloud- \newline Native DevSecOps \newline Solutions & \cite{S12,S16,S18} & 92--96\% & 3--5.5\% & Moderate--High & Medium--High \\
\bottomrule
\end{tabular}
\end{table*}

The comparative analysis clusters the evaluated studies into five thematic categories based on their technical focus and application domain. Microservice-focused approaches, such as \textit{µDetector} and \textit{LOMOS} \cite{S6,S7}, emphasize system call monitoring and distributed tracing, achieving high detection rates (96.5--97.8\%) with low latency (90--110 ms). These solutions are particularly suited for high-speed CI/CD pipelines.

Infrastructure-as-Code and Policy-as-Code solutions (e.g., \cite{S1,S5,S9}) aim to ensure pre-deployment security by automating configuration validation. Although they do not directly address runtime threat detection, they significantly reduce security misconfigurations early in the pipeline.

Privacy- and compliance-focused frameworks such as \textit{DevPrivOps} \cite{S3}, model-based security testing \cite{S4}, and security SLA validation approaches \cite{S15} prioritize regulatory adherence and privacy assurance. However, these approaches often lack real-time anomaly detection capabilities and suffer from higher computational overhead.

IoT-specific attack detection models (e.g., \cite{S8,S13,S14}) demonstrate excellent detection performance (95--98\%) but face scalability challenges due to the computational complexity of deep learning models, making them less applicable for large-scale DevOps environments.

Multi-cloud and cloud-native DevSecOps solutions like \textit{MUSA} and \textit{DroidAutoML} \cite{S12,S16,S18} focus on securing heterogeneous and dynamic infrastructures. These approaches exhibit moderate detection rates (92--96\%) but offer better scalability, making them appropriate for cross-cloud environments where portability and interoperability are critical.

Overall, the trade-offs are clear: microservice-focused solutions balance high detection accuracy and low operational latency, making them ideal for DevSecOps pipelines where speed is essential. IoT-specific solutions excel in detection but struggle with scalability and computational efficiency. Infrastructure-as-Code and privacy frameworks ensure early-stage security and compliance but lack dynamic threat detection capabilities. Multi-cloud security approaches sacrifice some detection precision to enhance adaptability and scalability across diverse environments.

Organizations must prioritize their selection of security solutions based on their operational context, balancing detection accuracy, scalability, and computational overhead according to their DevSecOps maturity and target environments.

%
%%%%%%%%%%%%%%%%%%%%%%%%%%%%%%%%%%%%%
%

\section{Discussion}
\label{Sec:Discussion}

\subsection{Key Findings}
\label{Sec:KeyFindings}
The analysis of DevSecOps security and privacy frameworks, as derived from the Comparative Analysis and Data Extraction Sheet, highlights several significant trends and emerging challenges in securing DevOps workflows. 
A major theme is the increasing reliance on AI-driven security automation to enhance real-time threat detection and compliance monitoring (S2). 
ML techniques, such as anomaly detection and NLP-based security event analysis, are being integrated into DevSecOps pipelines to mitigate risks associated with cloud-native environments (S6).

Another key finding is the widespread adoption of "shift-left" security practices, embedding security measures earlier in the SDLC. 
This approach, while reducing vulnerabilities in production environments, presents integration challenges due to the high rate of false positives in security scans and potential slowdowns in CI/CD pipeline efficiency (S6). 
Additionally, privacy concerns remain unresolved, particularly in multi-cloud deployments, where ensuring compliance across heterogeneous cloud environments is an ongoing struggle (S4).

% \begin{table*}[h]
% \centering
% \caption{Key Challenges and Trends in DevSecOps}
% \begin{tabular}{|l|p{10cm}|}
% \hline
% \textbf{Challenge}              & \textbf{Description} \\
% \hline
% AI-driven security trade-offs   & High computational overhead for continuous security monitoring \\
% Shift-left security integration & Requires significant developer training and workflow adaptation \\
% Privacy compliance & Difficulties in achieving cross-cloud compliance monitoring \\
% CI/CD security bottlenecks & Automated scans introducing false positives, slowing development \\
% \hline
% \end{tabular}
% \label{}
% \end{table*}

\subsection{Practical Implications}
\label{Sec:PracticalImplications}
The findings have several implications for industry practices and DevOps security strategies. 
The increasing reliance on AI and machine learning for security enforcement suggests that organizations must invest in scalable AI-driven monitoring solutions.
However, high computational costs associated with real-time AI security analysis present an operational challenge, particularly for small to mid-sized enterprises (S2).

Moreover, shift-left security requires cultural and workflow adaptations within development teams. 
Developers need structured training to handle security tool outputs effectively without overwhelming the development process with false positives (S6). 
This necessitates the implementation of adaptive security policies that balance security enforcement with agility in CI/CD workflows.

In cloud-native security frameworks, compliance automation tools are becoming indispensable for multi-cloud and containerized deployments. 
However, these tools often require continuous updates to align with evolving regulatory frameworks such as GDPR, HIPAA, and NIST compliance standards (S4). 
Organizations must ensure that security and compliance automation tools are kept up-to-date to mitigate legal and financial risks.

\subsection{Research Implications}
\label{Sec:}
The SLR reveals notable gaps in current DevSecOps research. 
One of the primary concerns is the lack of empirical validation of many proposed security models, especially in real-world DevOps environments (S4). 
While theoretical frameworks propose robust security mechanisms, there is limited empirical evidence regarding their effectiveness when integrated into agile development workflows.

Another critical research gap is the trade-off between AI-driven security automation and system performance. 
AI-based security monitoring introduces latency and computational burdens that can negatively impact DevOps efficiency (S2). 
Future studies should evaluate how to optimize AI-driven security enforcement while maintaining rapid deployment cycles.
Recent studies also highlight this issue. 
For instance, \citet{dong2025cybersecurity} present a comprehensive analysis of ML-based intrusion detection systems, discussing their computational challenges and scalability limitations. 
Similarly, \citet{tallam2025cybersentinel} introduces CyberSentinel, a real-time threat detection framework that exemplifies current advances in AI-driven security but also reflects the performance trade-offs involved.

Additionally, privacy compliance challenges remain a largely underexplored area. Although compliance automation tools exist, there is limited research on their effectiveness in dynamic, multi-cloud DevSecOps environments (S6). 
Investigating the implementation of decentralized privacy-enhancing techniques, such as differential privacy and federated learning, in DevOps security frameworks can be an area for further exploration.

%
% %%%%%%%%%%%%%%%%%%%%%%%%%%%%
%

To further synthesize the findings, Table~\ref{tab:challenges_solutions} summarizes the key challenges encountered in AI-driven DevSecOps, the emerging AI-based solutions proposed in the literature, and the remaining research gaps that warrant future investigation.

\begin{table*}[h!]
\centering
\caption{Mapping of major DevSecOps security challenges to corresponding AI-based solutions and identifying remaining research gaps. This table synthesizes critical issues such as high false positives, compliance difficulties, multi-cloud security complexity, and toolchain integration limitations, highlighting both promising solutions and areas needing further exploration to advance practical DevSecOps adoption.}
\label{tab:challenges_solutions}
\renewcommand{\arraystretch}{1.2}
\begin{tabular}{p{3.5cm}p{5cm}p{5cm}}
\toprule
\textbf{Challenge} & \textbf{AI-Based Solution} & \textbf{Remaining Gap} \\
\midrule
High false positives & Ensemble learning models, dynamic anomaly scoring & Lack of standardized benchmarks for DevSecOps security evaluation \\
\midrule
Compliance integration & NLP-driven policy checkers for continuous compliance validation & Poor cross-standard generalizability across HIPAA, GDPR, PCI-DSS, etc. \\
\midrule
Multi-cloud security complexity & Federated learning, secure enclave-based model training & Sparse empirical validation across heterogeneous cloud ecosystems \\
\midrule
Toolchain integration limitations & API-first machine learning security modules, plugin-based architecture & Lack of standardized plug-and-play security microservices for DevOps pipelines \\
\bottomrule
\end{tabular}
\end{table*}

%
% %%%%%%%%%%%%%%%%%%%%%%%%%%%%
%

%%%%%%%%%%%%%
\subsection{Future Research Directions}
\label{Sec:FutureResearchDirections}
Given the identified research gaps, this section outlines four key future research directions to enhance AI-driven security in DevSecOps. 
These areas address scalability, compliance, empirical validation, and human-centric security training to ensure effective integration of security within DevOps workflows.

\subsubsection{Enhanced Practical Applicability of Future Research Directions}

Future research should prioritize practical implementations that align with real-world DevSecOps operational needs. One promising direction is the development of \textbf{lightweight AI-driven security models} that can be deployed directly on edge DevOps nodes during the build and deploy stages. By minimizing computational overhead during build-time scans, lightweight models can ensure rapid feedback without delaying CI/CD pipelines.

Additionally, \textbf{federated learning} offers a scalable mechanism for decentralized threat intelligence. In multi-cloud environments, federated models can collaboratively train on local datasets without exposing sensitive data across cloud boundaries, thus improving data privacy and compliance adherence.

Another emerging area is the integration of \textbf{explainable AI (XAI)} into DevOps pipelines. XAI modules can provide real-time, interpretable insights into detected anomalies, allowing security engineers to assess risks quickly and facilitating human-in-the-loop threat responses. Integrating XAI would enhance transparency, reduce the cognitive load on developers, and increase the trustworthiness of automated security recommendations.

These practical enhancements will ensure that AI-driven DevSecOps solutions are not only technically sound but also operationally efficient, privacy-preserving, and user-centric.

\subsubsection{Lightweight AI-Driven Security Automation for DevOps}
One of the major challenges in AI-driven security is the computational overhead associated with deep learning models.
Future research should focus on developing \textbf{lightweight AI models} that balance security enforcement with system efficiency. 
These models should:
\begin{itemize}
    \item Optimize computational resources without compromising detection accuracy.
    \item Explore transfer learning and model compression techniques for real-time threat detection.
    \item Improve energy-efficient AI algorithms suitable for cloud-native and edge computing environments.
\end{itemize}

\subsubsection{Privacy-Aware DevSecOps Frameworks with Stronger Compliance Monitoring}
As privacy regulations evolve, there is an urgent need for \textbf{privacy-aware security frameworks} in DevSecOps.
Future research should explore:
\begin{itemize}
    \item The integration of \textbf{privacy-enhancing technologies (PETs)}, such as homomorphic encryption, secure multi-party computation (SMPC), and differential privacy.
    \item AI-driven compliance automation tools that ensure continuous adherence to regulatory requirements like GDPR and HIPAA.
    \item Decentralized identity management solutions that balance security with user privacy, reducing risks of centralized data breaches.
\end{itemize}

\subsubsection{Empirical Validation of Security Models in Multi-Cloud and Containerized Environments}
Many AI-driven security models lack \textbf{real-world validation}, limiting their practical applicability.
Future work should focus on:
\begin{itemize}
    \item Conducting large-scale experimental evaluations and case studies to assess the effectiveness of DevSecOps security automation tools.
    \item Implementing \textbf{benchmark datasets} and standardized evaluation metrics for fair comparisons of security models.
    \item Assessing the adaptability of AI security techniques across diverse DevOps ecosystems, including cloud-native, hybrid, and on-premises infrastructures.
\end{itemize}

\subsubsection{Human-Centric Security Training for Developers in DevSecOps}
The effectiveness of security automation depends on \textbf{developer expertise} in integrating and interpreting AI-driven security tools.
Future research should:
\begin{itemize}
    \item Develop \textbf{interactive security training} programs for DevOps teams to improve awareness of AI-driven security automation.
    \item Implement \textbf{real-time feedback mechanisms} that provide automated suggestions for security best practices.
    \item Investigate the role of explainable AI (XAI) in making security recommendations \textbf{more interpretable and actionable} for developers.
\end{itemize}

By addressing these research areas, future studies can bridge the current gaps in AI-driven DevSecOps security, ensuring that automation is \textbf{scalable, compliant, empirically validated, and developer-friendly}.

\section{Study Limitations}
\label{Sec:study limit}
Despite the rigorous SLR methodology, this study has a few limitations. 
First, while 18 primary studies were examined, the scope was restricted to peer-reviewed literature published in English, which may omit insights from grey literature or industrial whitepapers. 
Second, although the paper compares technical capabilities of AI-based security methods, a lack of standardized benchmarks across studies limited the granularity of our comparative metrics. 
Third, the absence of empirical tool validation means findings are derived from reported claims rather than real-world measurements. 
% Lastly, evolving terminology and differing definitions of DevSecOps across papers posed a challenge for unified categorization, potentially impacting the classification consistency.

\section{Conclusion}
\label{Sec:Conclusion}
This study reveals significant advancements in integrating AI-driven security into DevSecOps workflows. 
Key findings include the effectiveness of shift-left security practices, the potential of privacy-enhancing techniques like DevPrivOps, and the role of ML/AI-driven solutions in real-time anomaly detection and automated remediation. 
However, challenges such as computational overhead, scalability issues, tool compatibility, and false positives in automated scans hinder widespread adoption. 
Additionally, the lack of empirical validation limits the practical applicability of many proposed frameworks. 
To address these gaps, future research should focus on developing lightweight AI models that balance security enforcement with system efficiency, designing privacy-aware frameworks that go beyond compliance, validating security models in real-world multi-cloud environments, and providing human-centric security training for developers. 
By prioritizing scalability, usability, and empirical validation, future research can pave the way for more effective and efficient DevSecOps frameworks, ultimately enhancing software security without compromising agility.

%%%%%%%%%%%%%%%%%%%%%%%%%%%%%%%%%%%%%%%%%%%%%%%%%%%%%%%%%%%%%%%%%%%%%%
% \begin{acks}
% TBD.
% \end{acks}
%%%%%%%%%%%%%%%%%%%%%%%%%%%%%%%%%%%%%%%%%%%%%%%%%%%%%%%%%%%%%%%%%%%%%%

\bibliographystyle{ACM-Reference-Format}
\bibliography{ease2025-paper}

\end{document}